\documentclass[pra,twocolumn,aps,superscriptaddress]{revtex4-2}

\usepackage{hyperref}
\usepackage{graphicx}
\usepackage{xcolor}
\usepackage{amsfonts}
\usepackage{amsmath}
\usepackage{mathtools}
\usepackage{dcolumn}
\usepackage{amssymb}
\usepackage{physics}
\usepackage{sublabel}
\usepackage[utf8]{inputenc}

\usepackage{bm}

\begin{document}

\preprint{APS/123-QED}

\title{Optimized Current Density Reconstruction from Widefield Quantum Diamond Magnetic Field Maps}

\author{Siddhant Midha}
 \affiliation{Department of Electrical Engineering, Indian Institute of Technology Bombay, Powai, Mumbai--400076, India
}
\author{Madhur Parashar}
\affiliation{Department of Electrical Engineering, Indian Institute of Technology Bombay, Powai, Mumbai--400076, India
}
\author{Anuj Bathla}
 \affiliation{Center for Research in Nanotechnology and Science, Indian Institute of Technology Bombay, Powai, Mumbai--400076, India
}
\affiliation{Department of Electrical Engineering, Indian Institute of Technology Bombay, Powai, Mumbai--400076, India
}

\author{\\ David A. Broadway}
 \affiliation{School of Science, RMIT University, Melbourne, Victoria 3001, Australia
}
\author{Jean-Philippe Tetienne}
 \affiliation{School of Science, RMIT University, Melbourne, Victoria 3001, Australia
}
\author{Kasturi Saha}
\email{kasturis@ee.iitb.ac.in}
 \affiliation{Department of Electrical Engineering, Indian Institute of Technology Bombay, Powai, Mumbai--400076, India
}
\affiliation{Center of Excellence in Quantum Information, Computing Science and Technology, Indian Institute of Technology Bombay, Powai, Mumbai--400076, India}

\affiliation{Center of Excellence in Semiconductor Technologies (SemiX), Indian Institute of Technology Bombay, Powai, Mumbai--400076, India}

\date{\today}

\begin{abstract}
Quantum Diamond Microscopy using Nitrogen-Vacancy (NV) defects in diamond crystals has enabled the magnetic field imaging of a wide variety of nanoscale current profiles. Intimately linked with the imaging process is the problem of reconstructing the current density, which provides critical insight into the structure under study. This manifests as a non-trivial inverse problem of current reconstruction from noisy data, typically conducted via Fourier-based approaches. Learning algorithms and Bayesian methods have been proposed as novel alternatives for inference-based reconstructions. We study the applicability of Fourier-based and Bayesian methods for reconstructing two-dimensional current density maps from magnetic field images obtained from NV imaging. We discuss extensive numerical simulations to elucidate the performance of the reconstruction algorithms in various parameter regimes, and further validate our analysis via performing reconstructions on experimental data. Finally, we examine parameter regimes that favor specific reconstruction algorithms and provide an empirical approach for selecting regularization in Bayesian methods.
\end{abstract}

\maketitle

\section{Introduction}
Quantum-enabled magnetometry techniques provide invaluable insight into nanoscale structures by providing a spatial map of the magnetic field across the two-dimensional extent of the system. A variety of quantum systems have shown to be viable for the quantum sensing of magnetic fields, notable examples being superconducting quantum interference devices \cite{squid1, squid2, squid3}, neutral atoms \cite{neutral_atoms_1, neutral_atoms_2}, trapped ions \cite{trapped_ion_1, trapped_ion_2}, and solid-state systems such as nitrogen-vacancy (NV) centers in diamond \cite{imaging_nv_review, imaging_nv_wf, imaging_nv_principles}. Among these systems, NV centers stand out owing to their multi-modal ability to resolve changes in electromagnetic fields, temperature, and strain \cite{nv_multiple_sensor}, convenient room-temperature operation and optical readout \cite{nv_optical_readout}, and the ability to work as a single nanoscale sensor \cite{nv_single_sensor} or in wide-field-of-view ensemble sensing \cite{imaging_nv_wf}. 
This paper focuses on reconstructing current density using magnetometry facilitated by wide-field-of-view NV imaging or quantum diamond microscope (QDM), employing thin layers of NV centers embedded in bulk diamond crystals. This approach, particularly effective for large ensembles, has demonstrated per-pixel sensitivities reaching the $nT/\sqrt{Hz}$ level. Recently, applications such as imaging currents in micro-chip devices and integrated circuits \cite{imaging_nv_elecchip, nv_wf_microcirc, imaging_nv_current1, grapheneFET}, condensed matter systems \cite{imaging_nv_current2, imaging_nv_magn, shishir2021nitrogen} and biological systems \cite{nv_wf_bio,nv_wf_madhur} have been demonstrated.   \\\indent 
The magnetic field maps obtained using QDM are often used to compute the current density map, serving as a probe of the underlying physics in the sample under study. This reconstruction manifests as a non-trivial {inverse} problem. The forward problem of calculating the magnetic fields given the current density $\mathbf{j}$ across the sample is straightforward owing to the Biot-Savart law. On the other hand, the \textit{inverse} problem is not always well defined due to added noise and imperfections in measurement, and is ill-posed owing to direct inversions involving $1/0$ terms which require additional conditioning. This leads to the possibility of different noise conditions on the same sample resulting in non-unique reconstructed images. However, it has been shown that the inverse problem of computing the current density, given the magnetic field, is unique in the case of \textit{clean} magnetic field data produced by purely two-dimensional currents in the ideal case of infinite spatial resolution \cite{roth_using_1989}. But, even with a quasi-two-dimensional current distribution, \textit{noisy} magnetic field data superimposed with optical aberrations makes the inverse problem highly non-trivial and challenging. Owing to this non-uniqueness, many techniques have been developed to facilitate such inversions \cite{techniques_broadway_vector, techniques_Ghasemifard_sphericalharm, techniques_feldman_regul, techniques_kress, techniques_meltzer_g, techniques_tan, techniques_winjgaarden, roth_using_1989, clement_reconstruction_2021}. \\\indent 

\begin{figure*}[t]
\includegraphics[width=\textwidth]{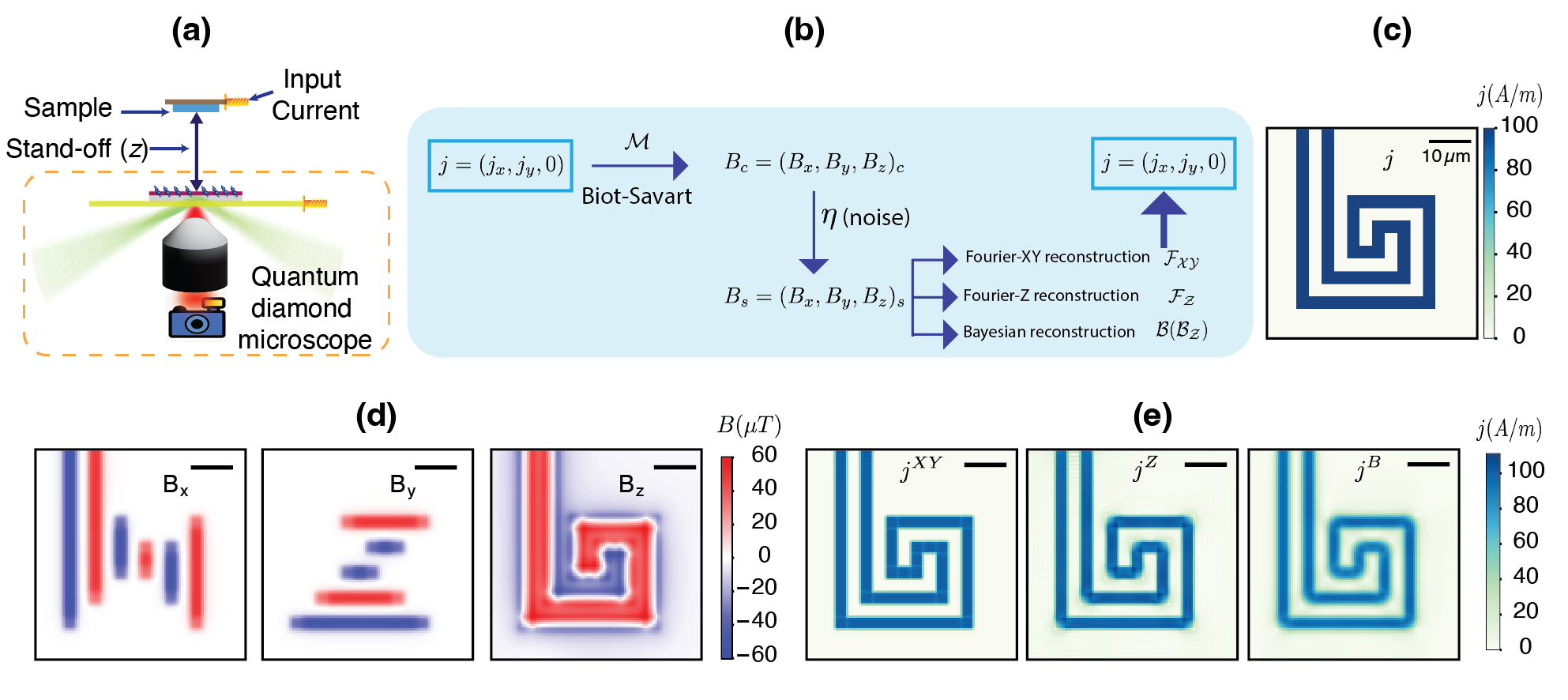}
\caption{(a) Schematic of quantum diamond microscope with a sample placed at a standoff distance denoted by $z$. (b) The block diagram of the simulated reconstruction protocol: the clean magnetic field $\mathbf{B}_c$ is computed from the current density $\mathbf{j}$ via the Biot-Savart kernel $\mathcal{M}$, followed by addition of noise $\eta$ to get the sensed fields $\mathbf{B}_s$. Subsequent post-processing steps using the Fourier-XY ($\mathcal{F}_{XY}$), the Fourier-Z ($\mathcal{F}_{Z}$) and the Bayesian ($\mathcal{B}$) methods result in different reconstructed current densities $\hat{\mathbf{j}}$. (c) Simulated spiral wire current density map. (d) Vector magnetic field maps computed from the current density without noise ($\eta = 0$) at a standoff of $z = 1\mu m$. (e) Reconstructed current density upon application of the three algorithms on the field data.}
\label{fig:fig1}
\centering
\end{figure*}

Traditionally, such inversion schemes deal with direct inversions in the Fourier space, as outlined by Roth et al. \cite{roth_using_1989} for the case of out-of-plane sensing. These Fourier-based methods have been expanded to encompass vector imaging, leading to enhanced reconstructions by utilizing all available magnetic field components \cite{techniques_broadway_vector}. Recent developments in the fields of machine learning and Bayesian inference have also led to the possibility of using Bayesian methods, where the aim is to learn the most probable, or `best fit' current density via maximization of the posterior probability $p(\mathbf{j}|\phi)$, where $\phi$ is the experimental magnetic field profile. This could be $\phi = B_z$ for the out-of-plane methods or $\phi = \{B_x, B_y, B_z\}$ for vector magnetometry. From an equivalent regularization point of view, such methods can be cast as optimization problems, where the aim is to reject the noise and retrieve the true underlying current density. At the outset, such Bayesian methods claim to offer the advantage of better reconstructions as prior knowledge about the sample can be imposed in the optimization process \cite{clement_reconstruction_2021, techniques_feldman_regul}. Nevertheless, there is lack of clarity on the applicability of such methods in comparison to vector and out-of-plane Fourier approaches, and no experimental advantage has been demonstrated thus far. \\\indent

In this work, we conduct an in-depth comparative analysis of the Fourier-based and Bayesian reconstruction methods for both simulated and experimental magnetic maps obtained from a quantum diamond microscope where the sample is assumed to be placed at a standoff distance $z$ from the source, as schematically outlined in Fig. \ref{fig:fig1}(a). The reconstruction pipeline is further outlined in Fig. \ref{fig:fig1}(b). We perform a thorough examination of the performance of reconstruction algorithms through extensive numerical simulations within various sensor regimes characterized by noise levels and standoff distances in Sec. \ref{sec:simresults}.  Following this, in Sec. \ref{sec:expresults} we build on our theoretical analysis by exploring practical approaches for determining the regularizer strength in Bayesian algorithms, and applying them to experimental data collected in three different noise and standoff scenarios. Through this, we thereby introduce the concept of Bayesian advantage in the realm of current density reconstructions for NV widefield experiments.


\section{Simulation results and discussions\label{sec:simresults}}
As a test structure, we use a microcoil wire---carrying a purely two-dimensional (2D) current to model a quasi 2D flow---with multiple sharp bends to effectively highlight nuances of reconstruction algorithms. A comprehensive outline of the simulation method, which includes the process for computing magnetic fields from current densities, as well as a detailed explanation of the reconstruction methodology is provided in the Supplementary Material \cite{SuppMat}. The assumed true current density, which by definition forms the ground truth is depicted in Fig. \ref{fig:fig1}(c), where we define the $j_x$ and $j_y$ components of the spiral wire piecewise in a total image size of $50 \times 50 \mu m^2$. The scale bar in Fig. \ref{fig:fig1} is $10 \mu m$. We assume a constant  current density of $100 A/m$ across the wire. The resulting magnetic field, computed at a standoff of $z = 1\mu m$, using a Fourier-space application of the Biot-Savart kernel $\mathcal{M}$, is shown in Fig. \ref{fig:fig1}(d).

\begin{figure*}[t]
\includegraphics[width=\textwidth]{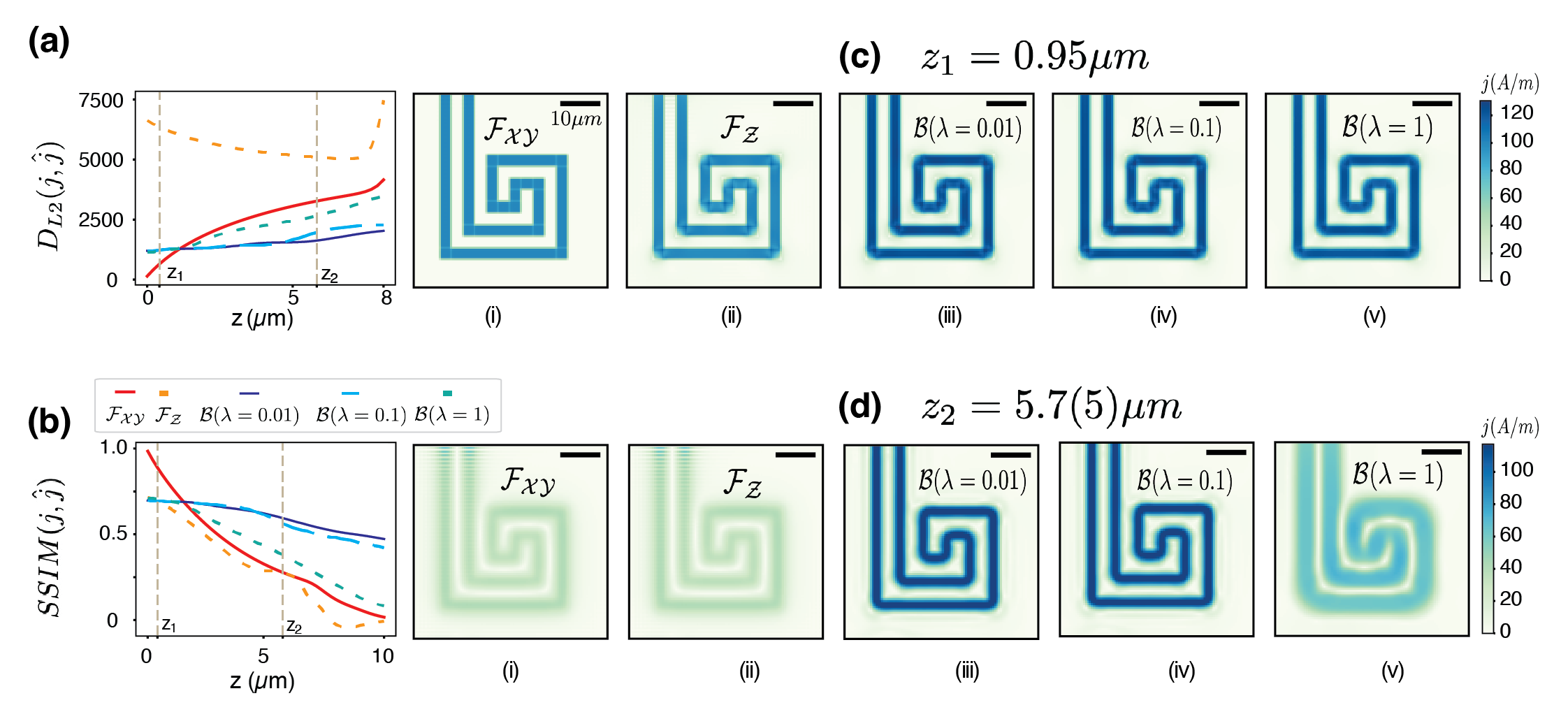}
\caption{Performance of the reconstruction algorithms as a function of the standoff distance in terms of the distance $D_{L2}$ and the structured similarity distance SSIM between the reconstructed and true current densities. (a) $D_{2L}$ as a function of standoff for $0 \leq z \leq 8 \mu m$. (b) SSIM as a function of standoff for $0 \leq z \leq 10 \mu m$. Two standoff points $z_1 = 0.95$ and $z_2=  5.7(5)$ (all in $\mu m$) are identified for further inspection. (c) Reconstructions at $z = z_1$.  (d) Reconstructions at $z = z_2$. The sub-panels of (b) and (d) show the reconstructions for Fourier-Z and the Fourier-XY methods (with a Fourier cutoff frequency of $\max\{2,\text{int}(z (\mu m))\}/z$) and the Bayesian reconstruction for three different regularizer strengths $\lambda \in \{0.01,0.1,1.0\}$.}
\label{fig:fig2}
\centering
\end{figure*}
\begin{figure*}[t]
\includegraphics[width=\textwidth]{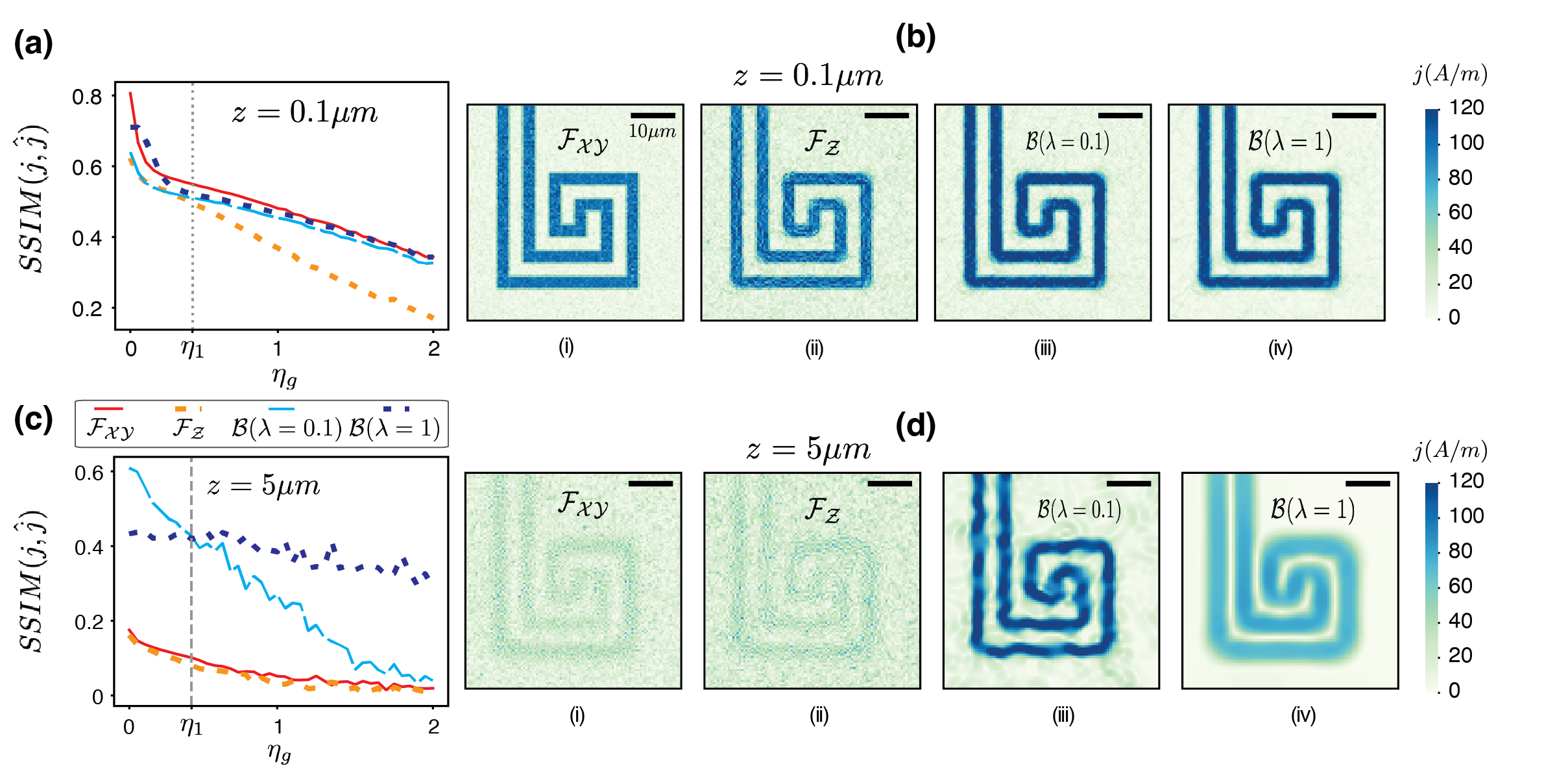}
\caption{Analyzing the effect of noise on the reconstructions using the SSIM as the error metric. (a-b) Results for the low standoff regime of $z =0.1\mu m$.
(c-d) Results for the high-standoff $z=5\mu m$ regime. The SSIM as a function of Gaussian noise $\eta_g$ is reported in (a) and (c) for the two regimes, while the panels (b) and (d) show the Fourier and Bayesian reconstructions at the $\eta_1 = 0.45$ noise level as marked with a dotted line. The Bayesian reconstructions are shown for two regularizer strengths $\lambda \in \{0.1,1.0\}$.}
\label{fig:fig3}
\centering
\end{figure*}

Reconstruction algorithms have access to one or more components of the magnetic field, as in Fig. \ref{fig:fig1}(d), and attempt to decode the underlying current density.  The Fourier-XY ($\mathcal{F}_{XY}$) method takes as input the $x$ and $y-$components of the magnetic field, and performs a Fourier space inversion for the current density. The Fourier-Z ($\mathcal{F}_Z$) performs this task with access only to the out-of-plane $z-$component. Lastly, we consider the Bayesian ($\mathcal{B}$) class of methods, which cast the inversion problem as inference, and solve it by performing Bayesian optimization. We consider a scenario in which the Bayesian algorithm has only the out-of-plane field component as input, emphasizing its effectiveness in situations with limited data, and simultaneously conducting a worst-case comparison with the Fourier-XY method. Intuitively, the task of finding the most probable current density can be converted into an optimization consisting of two terms: the first terms tries to fit to the data, and the second term attempts to create a smooth solution. The latter, termed a \textit{regularizer}, can be used to incorporate prior information about the current density into the optimization. A parameter $\lambda$ defines the strength of the regularizer, where a very small value $\lambda \to 0$ amounts to data fitting, and a very large value $\lambda \to \infty$ results in over-smoothened solutions. Thus, the regularizer strength $\lambda$ is a crucial part of the Bayesian reconstructions, and is used to ensure smooth and physical reconstructions.

The reconstructed $j$-maps, upon application of both the Fourier and Bayesian methods on the magnetic images in Fig. \ref{fig:fig1}(d) are shown in Fig. \ref{fig:fig1}(e). The regularizer strength $\lambda = \lambda^*$ is carefully selected in the Bayesian method, following a systematic and experimentally-informed procedure that we outline in the Supplementary Material \cite{SuppMat}. In the rest of the simulation results, we perform the comparative analysis of the Bayesian method in a non-adaptive setting \textit{without} performing regularizer selection, to outline worst case behavior. The following three sections outline the analysis of the three reconstruction algorithms on the simulated data in various noise and standoff regimes.  

\subsection{The Effect of Standoff}
The distance between the sample and the NV sensor, or the ``standoff" $(z)$, is a key parameter that strongly affects the measured magnetic field maps and consequently the underlying current distribution. This can be seen from the exponential dependence on standoff distance in the Biot-Savart kernel, $\mathcal{M}$ \cite{SuppMat}. This underscores the need for a thorough analysis of how the performance of the algorithms is influenced by the standoff distance.

We plot two error-estimation metrics, the two-normed distance ($D_{L2}$) and structural similarity index (SSIM), first introduced in \cite{wang2004image}, as a function of standoff $z$ in Fig. \ref{fig:fig2}(a) and (b) respectively, without any added noise ($\eta = 0$). See \cite{similarity_review,error_metrics_review} for brief reviews of the error and similarity metrics. The two-normed distance measures the Euclidean distance between two images, and the SSIM measures a `score' of visual similarity \cite{SuppMat}. The Bayesian reconstructions require selecting the strength of \textit{regularizer} term, which maintains smoothness of the reconstructed solution. Here, we do not perform systematic regularizer selection and rather choose three fiducial strengths  $\lambda \in \{0.01, 0.1,1\}$ in a \textit{non-adaptive} fashion. The Fourier methods on the other hand require choosing a cut-off frequency to band-limit the spectrum and perform inversions \cite{SuppMat}. We perform adaptive Fourier reconstructions ($\mathcal{F}_{XY}$, $\mathcal{F}_{Z}$) with the cutoff frequency of the algorithms changing with every $2\mu m$ increase in standoff, as elaborated in \cite{SuppMat}. Two points representing the low standoff $z_1 \lesssim 1\mu m$ and high standoff $z_2 \gtrsim 5\mu m$ are marked in the error plots with dotted lines. 
First, we observe that in both $D_{2L}$ and SSIM, the Fourier-XY method ($\mathcal{F}_{XY}$) performs better for the low-standoff regime of $z \lesssim 1\mu m$. As the standoff is increased, there exists a critical point beyond which the Bayesian method ($\mathcal{B}$) method starts to outperform $\mathcal{F}_{XY}$. 
This can be explained by noting that the $\mathcal{F}_{XY}$ method is exact at low stand-off, eliminating aberrations due to its ability to access both in-plane components, however, the spiral wire contains a non-trivial frequency spectrum owing to multiple sharp corners, and thereby requires a careful selection of a band-limiting frequency for the Fourier algorithms at higher standoff. Additionally, without such a selection of the cutoff frequency in the Fourier algorithms at higher standoffs, high-frequency artifacts dominate the Fourier reconstruction. This leads to a sharp unbounded increase in the $D_{2L}$ metric at $z \approx 8\mu m$. In contrast, the Bayesian reconstruction exhibits a SSIM decrease less than $10\%$ for standoff distances of $z \lesssim 5\mu m$.

Fig. \ref{fig:fig2}(c) and (d) show reconstructed current densities at $z=0.95\mu m$ and $z=5.7(5)\mu m$ standoff respectively. At $z = z_1$ in Fig. \ref{fig:fig2}(c), none of the reconstructions show any artifacts, and resolve the widths of all the arms correctly. The Bayesian images show an offset above the true value, which is attributed to the fact that the regularizer strengths are chosen arbitrarily. In Fig. \ref{fig:fig2}(d), when the standoff is increased, noticeable discrepancies in the Fourier images become apparent, including an offset and a notable decline in resolution. However, the Bayesian image, particularly with $\lambda \lesssim 0.1$, exhibits a significantly improved outcome. It is instructive to note that the Bayesian image with $\lambda = 1$ is significantly broadened which is attributed to the over-smoothening in the optimization process owing to a larger regularizer strength than required. Both the Fourier and Bayesian algorithms necessitate the fine-tuning of parameters within a fixed set of fields $\phi$, either through the specification of a band-limiting frequency or the adjustment of a regularizer strength. It is crucial to highlight that achieving precise fine-tuning in the Fourier method becomes challenging at higher standoffs, particularly when dealing with non-trivial frequency spectra, and no systematic method exists for such tuning in the Fourier methods. Additionally, the results presented in Fig. \ref{fig:fig2} indicate the strict requirement for tuning of parameters for the Fourier method as the standoff increases, contrasting with the Bayesian method, which requires minimal tuning and $\mathcal{B}(\lambda \lesssim 0.1)$ consistently yields high-quality reconstructions as the standoff increases. It is noteworthy that while the $D_{2L}$ in Fig. \ref{fig:fig2}(a) method exhibits a substantial difference in the performance of $\mathcal{F}_{XY}$ and $\mathcal{F}_{Z}$, particularly around $z \approx z_1$, the reconstructions depicted in Fig. \ref{fig:fig2}(c) showcase similar performance of the two methods. The structural similarity index effectively reflects this observation, emphasizing that SSIM distance provides more perceptual information regarding the algorithms' performance compared to the two-normed distance. Consequently, SSIM, being constrained by the condition SSIM $\leq 1$, is considered a superior metric. In the rest of the paper, we use SSIM exclusively for quantifying the error, resulting in a bounded, well-informed estimation metric.

\subsection{Robustness to Noise}
We now study the effect of noise on reconstruction. We employ two noise models---Gaussian noise, which distributes across the entire image uniformly, and \textit{scaled noise}, which is more prominent at pixels with higher absolute values. Both the models add zero-mean Gaussian noise to each pixel, with the former having a pixel independent variance and the latter having a variance that scales with the absolute value of the pixel. The amount of noise in both models is controlled by a single parameter $\eta$ \cite{SuppMat}. We study the SSIM distance between the reconstructed and ground truth current densities as a function of noise strength $\eta$. Results are shown here for Gaussian noise, with scaled noise exhibiting similar patterns---differences between the two noise models are shown in the next subsection. Figure \ref{fig:fig3} shows the results for (a-b) the low-standoff regime of $z=0.1\mu m$ and (c-d) the high standoff regime of $z=5\mu m$. For Bayesian reconstructions, $\lambda \in \{0.1,1.0\}$ is chosen. The $\lambda = 0.01$ case is omitted as the presence of noise warrants non-trivial regularization. In the scenario with a low standoff, the $\mathcal{F}_{XY}$ method emerges as the most effective, displaying the highest SSIM score for the entire noise range $0 \leq \eta_g \leq 2$. The SSIM also reveals two distinct patterns at low standoff --- a rapid decline in algorithm performance with increasing noise, followed by a shift in slope after reaching a critical noise strength. This phenomenon can be linked to a critical signal-to-noise (SNR) ratio, where the noise levels equate to the signal strength, resulting in two performance regimes. The $\mathcal{F}_{Z}$ image exhibits a low SNR ratio within the wire, leading to diminished SSIM as noise levels escalate. In contrast, the $\mathcal{B}(\lambda)$ reconstructions exhibit high SNR, as well as the characteristic offset associated with the lack of regularizer selection.

The outcomes for the high-standoff regime are depicted in Fig. \ref{fig:fig3}(c-d). The SSIM plot indicates a consistent score (SSIM $\approx 0.4$) performance for the $\mathcal{B}(\lambda = 1)$ reconstructions, and reveals intriguing behavior for $\mathcal{B}(\lambda = 0.1)$. The $\lambda = 0.1$ approach surpasses all other reconstructions until reaching $\eta_1 = 0.45$, after which it exhibits a score lower than the $\lambda = 1$ reconstruction. This phenomenon can be explained as follows: at minimal noise levels, the appropriate regularizer strength is closer to $\lambda \sim 0.1$, resulting in accurate resolution and noise rejection. Subsequently, in this high standoff scenario, with increasing noise, there is a swift change in the required regularizer strength, leading to a rapid decline in performance. This analysis is extended by presenting the reconstructed densities at $\eta = \eta_1$ in Fig. \ref{fig:fig3}(c) for this case. The $\mathcal{B}(\lambda = 0.1)$ reconstruction correctly resolves the wire but exhibits substantial spurious noise outside the wire, as anticipated due to the crossover at $\eta = \eta_1$. In other words, the consistent performance of the $\lambda = 1$ method stems from the fact that the regularizer strength is much larger than the optimal value, resulting in effective noise rejection but significant broadening overall. The optimal value of the regularizer lies between the two extremes depicted. On the other hand, the Fourier $(\mathcal{F}_{XY}, \mathcal{F}_{Z}$) reconstructions show a very small and decreasing similarity score SSIM $\lesssim 0.2$, as can be seen from the reconstructions in Fig. \ref{fig:fig3}(d).

Therefore, we deduce that under conditions of very low standoffs and moderate noise, the $\mathcal{F}_{XY}$ approach stands out as a promising option for widefield reconstructions when provided with access to all in-plane measurement data. In scenarios where only out-of-plane data is available, the $\mathcal{B}$ method demonstrates comparable or superior performance compared to $\mathcal{F}_Z$ at low standoffs. As standoff distances increase, the utilization of Bayesian techniques becomes imperative, with the $\mathcal{B}$ method consistently outperforming $\mathcal{F}_Z$ across all noise levels, even without the need for regularizer tuning for moderate noise levels $\eta \lesssim \eta_1$. It is important to note that the distinction between low and high standoff scenarios is representative and may be influenced by factors such as wire length-scales relative to image length and the amount of current in the wire, impacting SNR. The selection of a high current density magnitude of $100 A/m$ enables to delineate these two regimes at $z \in \{0.1,5\} \mu m$, which can be realized by varying the standoff in experimental settings.

\subsection{Phase Diagram}
\begin{figure*}[t]
\includegraphics[width=\textwidth, height = 5cm]{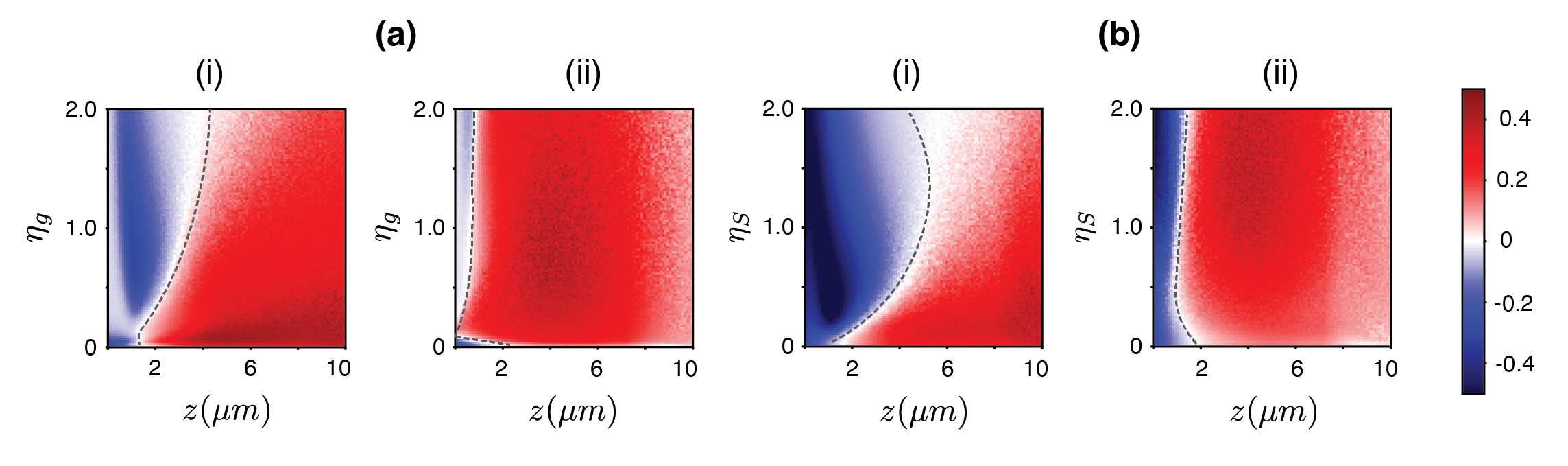}
\caption{Phase diagram of Bayesian advantage. The SSIM metric for the Bayesian and Fourier-XY reconstructions with the true densities is computed as a function of the standoff $z$ and the noise level $\eta$, and their difference is shown. (a) Results for Gaussian noise. (b) Results for scaled noise. Sub-panels are for regularizer strengths (i) $\lambda = 0.1$ and (ii) $\lambda = 1.0$. The red and blue regions represent Bayesian and Fourier advantage, respectively, and the dotted line marks the phase transition boundary.}
\label{fig:fig4}
\centering
\end{figure*}

We now investigate the combined effects of $\eta$ and $z$ to define and demonstrate \textit{Bayesian advantage}. The standoff and the noise is varied in the range of $z \in [0,10] \mu m$, and $\eta_{g/s} \in [0,2]$ respectively. To quantify Bayesian advantage, we construct a map of the SSIM distance as a function of $\eta$ and $z$ for the $\mathcal{B}$ and $\mathcal{F}_{XY}$ methods and define,
\begin{equation}
    d(z,\eta) := \text{SSIM}(z,\eta; \mathcal{B}) -  \text{SSIM}(z,\eta; \mathcal{F}_{XY})
\end{equation}
where $ \text{SSIM}(z,\eta; \mathcal{X})$ denotes the SSIM computed in the case of standoff $z$, noise $\eta$ and the algorithm $\mathcal{X}\in\{\mathcal{B},\mathcal{F}_{XY}\}$. Note that $-1 \leq d(z,\eta) \leq 1$, where $d > 0$ and $d < 0 $ indicate Bayesian and Fourier-XY advantage, respectively. The Fourier-XY method is chosen as a comparative candidate of Bayesian advantage owing to its superior performance in the Fourier methods, and parameters for both the algorithms are held constant.

The outcomes for $d(z,\eta)$ are presented in Fig. \ref{fig:fig4} for (a) Gaussian and (b) scaled noise, with sub-panels illustrating the model selection for the Bayesian approach using $\lambda \in \{0.1,1\}$ in accordance with the earlier discussions. The dashed line signifies a distinctive \textit{phase transition} (in the spirit of similar phase diagrams in optimization dynamics, for instance see \cite{kalra2023phase}) from Fourier to Bayesian advantage, for which $d(z,\eta)$ serves as an effective order parameter. In the case of Gaussian noise in Fig. \ref{fig:fig4}(a), results indicate that at very low standoffs, there is no clear preference for either of the two methods, as was seen in Fig. \ref{fig:fig3}(a). Additionally, for $\lambda = 1$ in the Bayesian scenario, the Fourier advantage region diminishes, suggesting that adequate regularization in Bayesian optimization generally leads to improved reconstruction quality without the need for regularizer tuning. In the case of scaled noise in Fig. \ref{fig:fig4}(b), similar to a `phase transition,' the results show a notable contrast with a pronounced Fourier advantage at low standoffs, as evidenced by the image contrast. Moreover, the area of Fourier advantage is smaller in the Gaussian noise case (Fig. \ref{fig:fig4}(a)) as compared to the case of scaled noise (Fig. \ref{fig:fig4}(b)). This is because of the fact that Gaussian noise has a flat and non-decaying spectrum for any $\eta_g > 0$, playing a more detrimental effect in Fourier inversions.

Given the consistent Bayesian advantage in the presence of Gaussian noise, theoretically, employing an adaptive regularizer scheme would yield a $d(z,\eta)$ plot that is consistently positive. Furthermore, these findings indicate that a significant portion of the experimental scenarios (quantified here as the $\eta-z$ plane) demonstrate the Bayesian advantage. This necessitates a thorough discussion of such Bayesian protocols on experimental data, which is conducted in the following section.

\begin{figure}[t]
\includegraphics[width=0.5\textwidth]{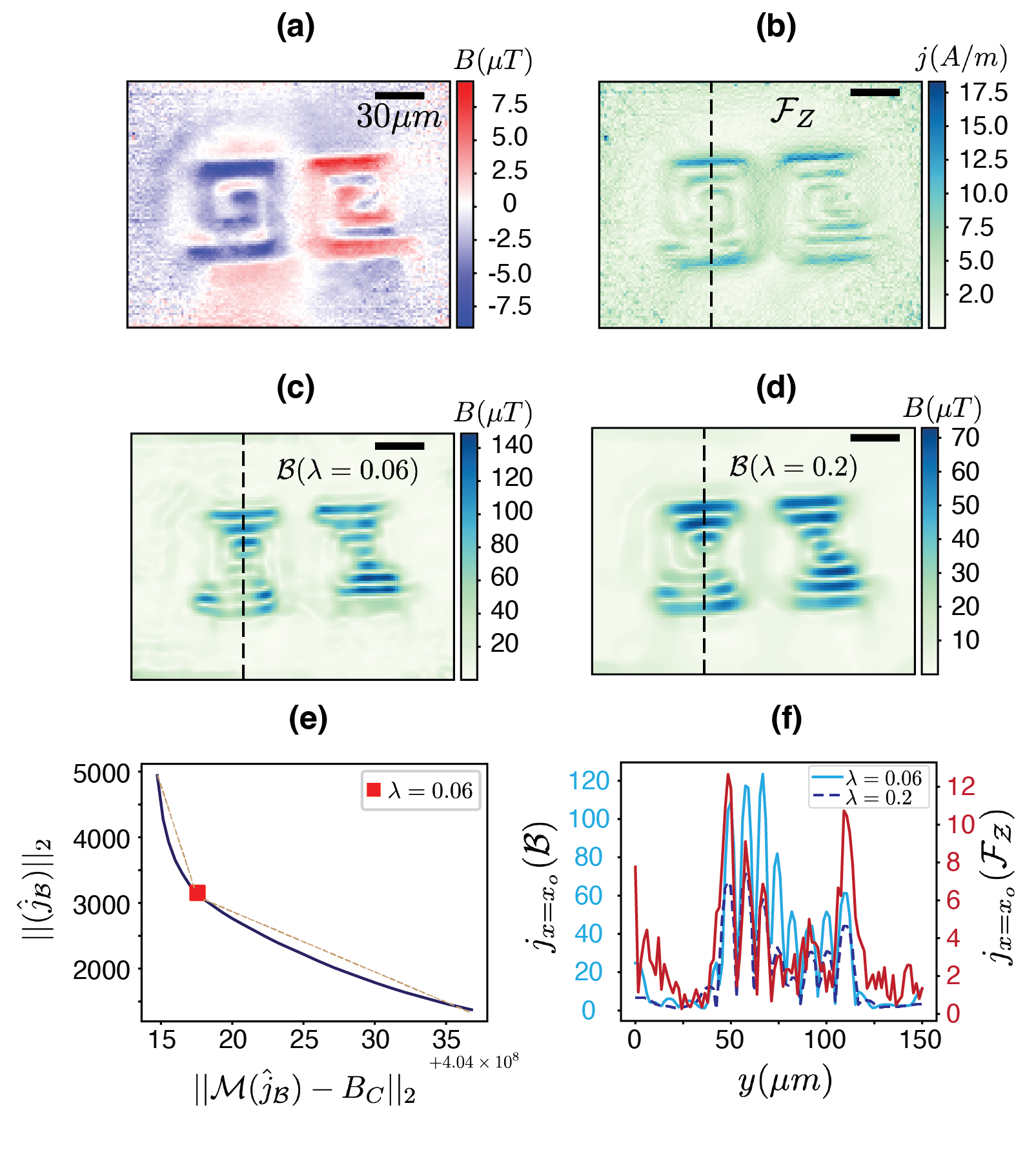}
\caption{Analysis of the micro-spiral wire. (a) Experimentally acquired magnetic field map projected along a given NV axis, with an estimated standoff of $z\sim 5\mu m$. (b) Reconstruction performed using the Fourier-Z ($\mathcal{F}_Z$) method. (c-d) Bayesian reconstructions for two different regularizer strengths $\lambda \in \{0.06, 0.2\}$ respectively. Dotted line marks the point $x_0 = 68\mu m$ in the reconstructions. (e) $l$-curve showing a turn at $\lambda = 0.06$. (f) Line-cut plots along the dotted line at $x = x_0$. Scale bars are $30 \mu m$.}\label{fig:fig1exp}
\centering
\end{figure}

\begin{figure*}[t]
\includegraphics[width=1\textwidth]{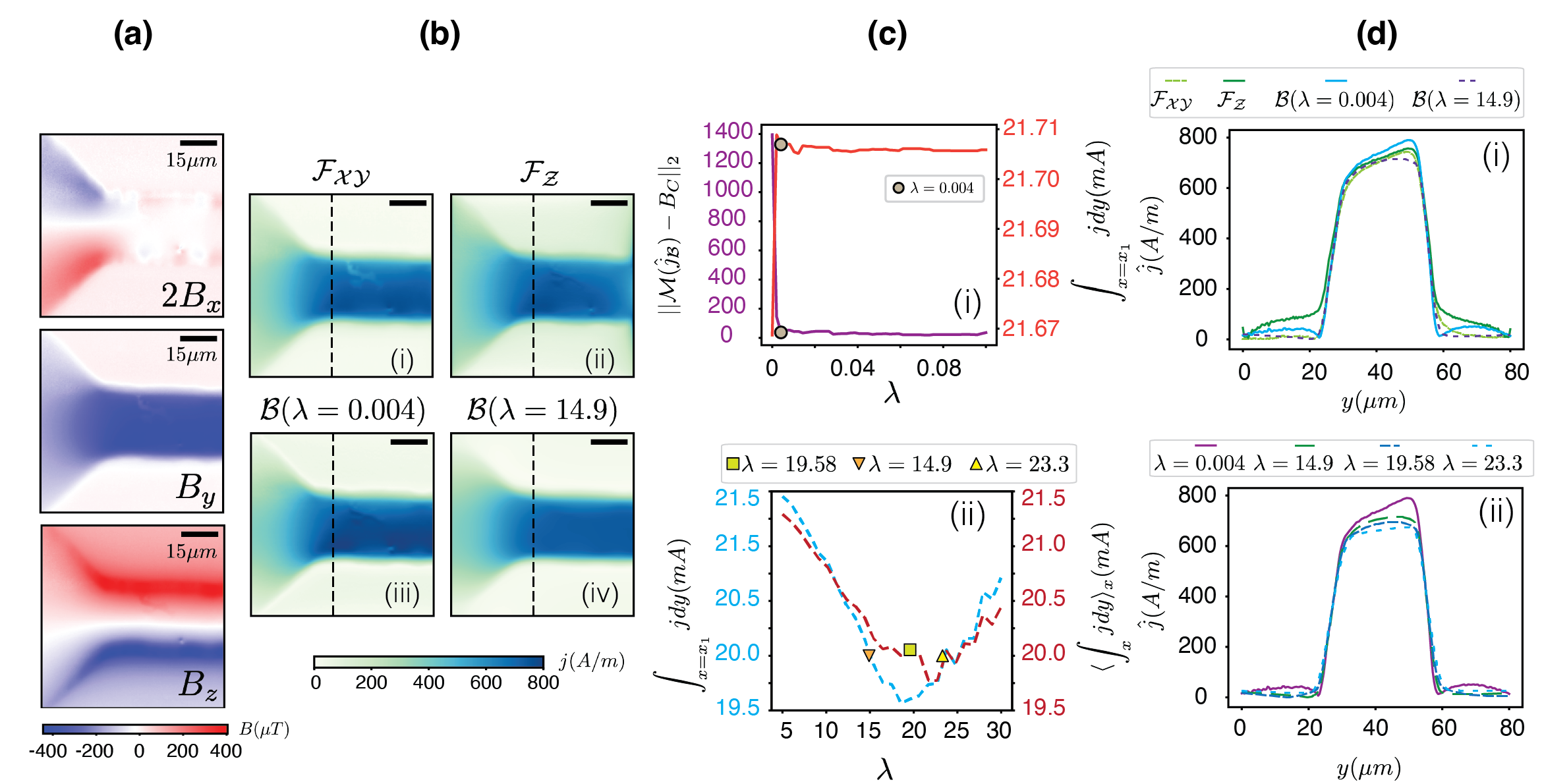}
\caption{Analysis of the niobium wire. (a) Magnetic field data obtained from NV widefield imaging on the Nb-wire, with a standoff $z \sim 0.1\mu m$. Data is taken from Ref. \cite{techniques_broadway_vector}. (b) Reconstructed current densities using the Fourier-XY and Fourier-Z methods, and Bayesian reconstructions for $\lambda \in \{0.004, 14.9\}$ with a dotted marked line at $x_1=36\mu m$. (c)(i) The magnetic field two-norm distance (left axis) and the total integrated current at $x=x_1$ (right axis) for small regularizer strengths $0 \leq \lambda \leq 0.1$. (c)(ii) The current at $x=x_1$ (left axis) and the total averaged integrated current (right axis) for large regularizer strengths $5 \leq  \lambda \leq 30$.  (d)(i) Line-cut plots of the density maps for the Fourier and Bayesian $\lambda \in \{0.004,14.9\}$. (d)(ii) Line-cuts for Bayesian reconstructions with different regularizer strengths $\lambda \in \{0.004,14.9,19.5(8),23.3\}$. Scale bars are $15 \mu m$.}
\label{fig:fig2exp}
\centering
\end{figure*}

\begin{figure*}[t]
\includegraphics[width=1\textwidth]{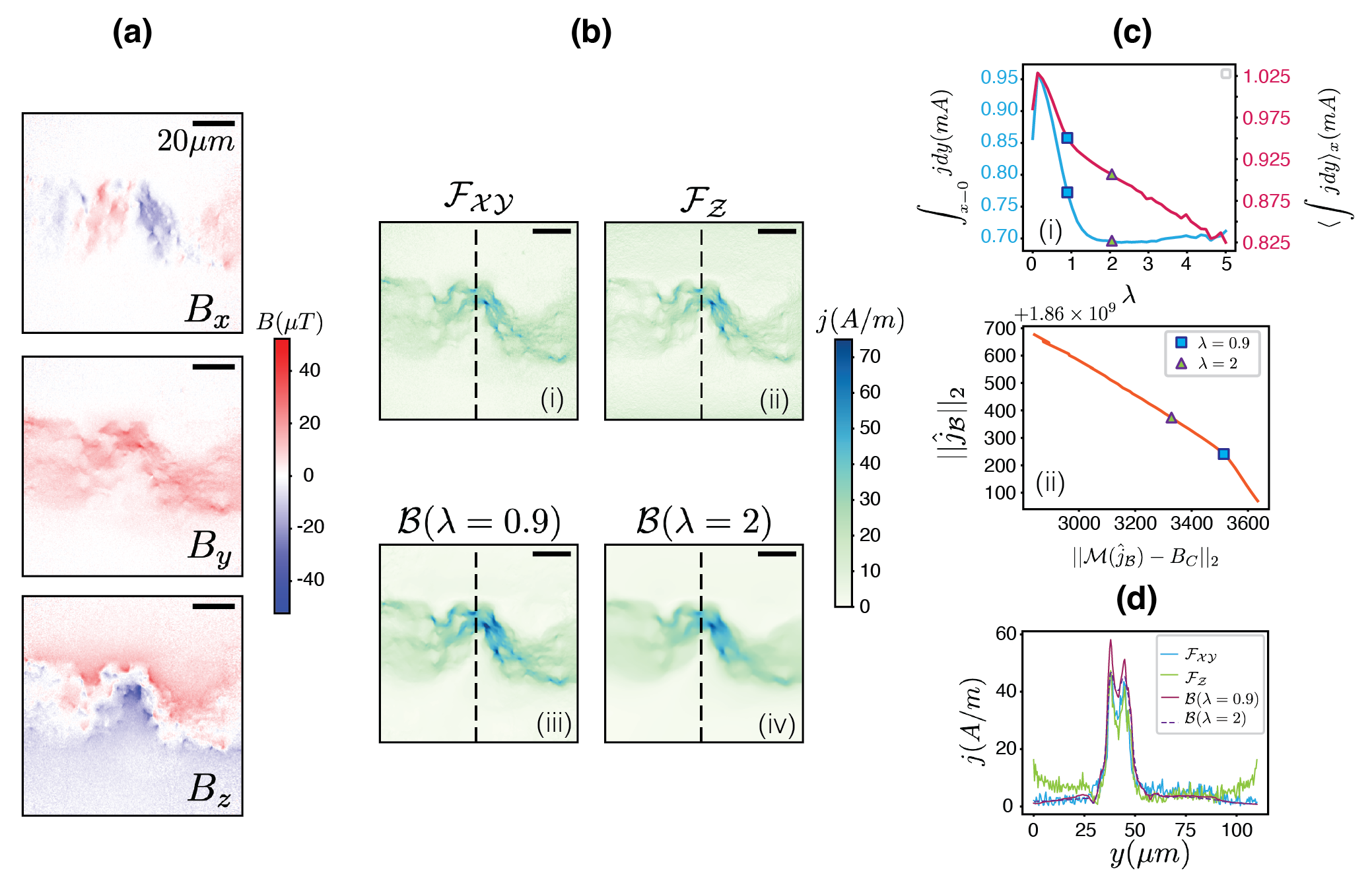}
\caption{Analysis of the graphene wire. (a) Magnetic field data obtained from NV widefield imaging on the graphene device, with a standoff $z \sim 20 nm$. Data is taken from Ref.~\cite{grapheneFET}. (b) Reconstructed density magnitudes using the Fourier and the Bayesian method by using $\lambda\in\{0.9,2\}$ with $x_2 = 54\mu m$ marked with a dotted line. (c)(i) Total integrated current entering from the left at $x=0$ (left axis) and average value of the total integrated currents (right axis) with $\lambda\in\{0.9,2\}$ marked. (c)(ii) $l$-curve showing a turn at $\lambda = 0.9$. (d) Linecut plots along the dotted line at $x=x_2$. All scale bars are $20\mu m$.} \label{fig:fig3exp}
\centering
\end{figure*}

\section{Reconstruction from Experimental Magnetic Maps \label{sec:expresults}}

In the previous discussions, we demonstrated how inference-based protocols exhibit an advantage in current density reconstruction by studying simple uncorrelated noise. In this section, we further show that the same models indeed work on experimental data, presumably incurring non-analytical and correlated noise, and further exemplify the Bayesian advantage. We now apply the reconstruction algorithms to three chosen experimental scenarios to illustrate the points described via simulations in the previous section. The regularizer selection---the search of the optimal trade-off $\lambda^*$---for Bayesian reconstructions on the experimental data is performed using the $l$-curve method and via an analysis of the total integrated currents. The $l-$curve method involves searching for a bending point in a parametrized plot of the norm of the reconstructed density $||\hat{j}_{\mathcal{B}}||_2$ against the norm of the residual $||\mathcal{M}(\hat{j}_{\mathcal{B}}) - B_c||_2$ as the $\lambda$ parameter is varied  {(see \cite{hansen1999curve,Kindermann_2020})}. The key idea behind selecting a near-optimal regularizer is to include correct prior information about the expected current densities in the optimization process. Mathematically, we inforce a particular functional form, known as the total-variation regularizer \cite{TV_lcurve,TVCV} which ensures smootheness of the reconstruction by penalizing the first derivatives. The strength of this prior is then controlled by the parameter $\lambda$. A small $\lambda\approx 0$ fits to the data, which can include the noise, and will produce exactly the same results as the Fourier inversion in the case of no noise ($\eta = 0$). The ingenuity of the Bayesian methods lies in tuning the $\lambda$, to be able to control the tradeoff between fitting the exact noisy data and rejecting the noise to reconstruct the correct fields. Details of the regularizer selection method are outlined with the simulated micro-coil wire as an instructive example in the Supplementary Material \cite{SuppMat}.

\subsection{Micro-spiral wire}
The micro-spiral wire consists of two spiral coils placed next to each other, carrying currents of opposite polarity. The magnetic field image of the micro-spiral obtained through NV widefield imaging is shown in Fig. \ref{fig:fig1exp}(a). We note that a lack of contrast at the left and right-most arms of the wire is present owing to imperfections in the imaging. We reconstruct the current density maps via $\mathcal{F}_{Z}$ and $\mathcal{B}$ using the out-of-plane $B_z$ component of the magnetic field.

The Fourier reconstruction is displayed in Fig. \ref{fig:fig1exp}(b), and the Bayesian reconstructions for regularizer strengths $\lambda\in\{0.06,0.2\}$ are depicted in (c) and (d) respectively. A dashed line is positioned at $x_0 = 68\mu m$ for further investigation through line cuts. The selection of $\lambda = 0.06$ is determined from the $l$-curve illustrated in (e), where the turning point is identified by the asymptotic lines drawn. The challenging balance between data-fitting and noise-rejection, given the wire's current of $500\mu A$, a high standoff of $\sim 5\mu m$, and noise in the image leads to a clear distinction between two asymptotic regimes.

Upon closer inspection, the Fourier image reveals numerous spurious currents outside the wire region, reaching up to $50\%$ of the maximum value. Additionally, there is a noticeable loss of resolution in the inner arms of the wire in the Fourier image, particularly along the $x= 68\mu m$ dotted line. Furthermore, the magnitude of the Fourier image is inaccurate, resulting in a total integrated current of $503(4) \mu A$ at the $x=68\mu m$ line, significantly deviating from the expected value of $ 8 \times 500 \sim 4000 \mu A$ due to the wire's eight arms. In contrast, the Bayesian reconstruction for $\lambda = 0.06$ exhibits precise resolution of the inner arms, with the left and right-most arms remaining unresolved. The total integrated current at $x=68\mu m$ is $4320(6) \mu A$, aligning with the expected $4000\mu A$ value. With a larger strength of $\lambda = 0.2$, fewer noise artifacts are observed outside the wire, but the wire's width increases, resulting in a total current of $2661(5)\mu A$. However, due to increased regularization, this reconstruction offers improved resolution of the left and right-most arms.

Line-cut plots along the dashed line of the current maps at the position $x = 68\mu m$ elucidate these observations. The $\mathcal{B}$ cuts exhibit sharper resolution of the arms and less noise outside the wire region. The resolution of the $\mathcal{B}(\lambda = 0.06)$ line-cut is particularly evident, while the $\mathcal{F}_Z$ plot displays reduced resolution of the arm at $y=100\mu m$ and an inaccurate magnitude of the density. Given the noise and standoff conditions, the $\mathcal{B}(\lambda = 0.06)$ reconstruction emerges as the most favorable in the micro-spiral wire.

\subsection{Niobium wire}

In contrast to the micro-spiral wire with large noise and standoff, we now investigate the reconstruction algorithms on a Nb-wire \cite{techniques_broadway_vector}, as shown in Fig. \ref{fig:fig2exp}, carrying a high total current of $20mA$ leading to a high SNR, with a low standoff $z \sim 0.1\mu m$ and a large wire-length to image-length ratio. This experiment represents the polar opposite case of the micro-spiral wire. According to the simulations and as per the phase diagram analysis, we expect no Bayesian advantage in this experiment. The vector magnetic field maps are shown in the three panels of Fig. \ref{fig:fig2exp}(a). We observe that in the $B_z$ map the field at the top and bottom edges of the image is not fully decayed, being still around $25\%$ of the maximum values, which is expected to cause artificial edge currents in the $\mathcal{F}_Z$ reconstructions due to data truncation. We also note the presence of a physical defect in the magnetic images in a region around the point $x=y=45\mu m$ (from the bottom-left corner).

The reconstructions utilizing the $\mathcal{F}_{XY}$ and $\mathcal{F}_Z$ methods are depicted in Fig. \ref{fig:fig2exp}(b) (i-ii) respectively. The $\mathcal{F}_{XY}$ illustrates an immaculate reconstruction, featuring a distinct identification of the defect in the fields, absence of noise artifacts, and a total integrated current of $20mA$ at $x=x_1$ (indicated by the dotted line in the reconstructions). In contrast, the $\mathcal{F}_{Z}$ image is derived after subtracting bias to address artifacts induced by truncation effects, albeit still exhibiting some edge artifacts towards the right of the image. The total current in this case is $21(8)mA$, deviating from the expected value of $20mA$.

The Bayesian reconstructions are displayed in Fig. \ref{fig:fig2exp}(b) (iii-iv) for the regularizer strengths $\lambda \in \{0.004,14.9\}$ respectively. The selection of $\lambda = 0.004$ is determined from a plot of the residual norm $||\mathcal{M}(\hat{j}_{\mathcal{B}}) - B||_2$ and the total current at $x = x_1$ as functions of $\lambda$, illustrated in Fig. \ref{fig:fig2exp}(c)(i) for small regularizer strengths $\lambda \leq 0.1$. With an increase in regularizer strength, critical points emerge in both the total current at $x=x_1$ and the averaged total currents, as depicted in Fig. \ref{fig:fig2exp}(c)(ii) for $\lambda $. Owing to the absence of sufficient noise to indicate a turning point in the $l$-curve, it is not shown here. The significant $\lambda$ points in $\{14.9,19.58,23.3\}$ exhibit similar reconstruction features; thus, the reconstructed image for $\lambda = 14.9$ is presented, where both the average current and the current at $x =x_1$ align at $\sim 20mA$. It is noteworthy that the $\mathcal{B}(\lambda = 0.004)$ reconstruction reveals sharper underlying features, such as the defect, reporting a current of $21(7)mA$ at $x_1$. In contrast, the $\mathcal{B}(\lambda = 14.9)$ image, influenced by the large regularizer, averages out finer features but reports a current of $20(1)mA$ at $x=x_1$. This emphasizes that the $l$-curve method is not always applicable in experimental scenarios where the fitting and smootheneing regimes are difficult to delineate, and thus an analysis of the total integrated current and residual magnetic field errors must be conducted in order to select $\lambda \approx \lambda ^*$. The line-cut plot in Fig. \ref{fig:fig2exp}(i) elucidates this point, and Fig. \ref{fig:fig2exp}(ii) shows the subtle difference between the Bayesian reconstructions for different regularizer strengths. While the $\mathcal{B}$ reconstructions show no edge artifacts and are superior to the $\mathcal{F}_Z$ images, the $\mathcal{F}_{XY}$ method emerges superior owing to this difficulty in the Bayesian trade-off. 

\subsection{Graphene wire}

Finally, we apply the reconstruction algorithms on imaging of a graphene wire using widefield NV imaging \cite{grapheneFET}. The standoff is again low ($z \sim 20 nm$) but the current pattern is much more complex than in the Nb-wire case. The results are shown in Fig. \ref{fig:fig3exp}. The vector magnetic field maps are shown in the three panels of Fig. \ref{fig:fig3exp}(a). Before discussing the results, we note that the total current flowing from the source (left) into the graphene channel is $700\mu A$ which, along with the fine features of the graphene wire, implies a low SNR in the image.

The reconstructions for the $\mathcal{F}_{XY}$ and $\mathcal{F}_Z$ cases are depicted in Fig. \ref{fig:fig3exp}(b)(i-ii), while the outcomes for the $\mathcal{B}$ algorithm, employing two different regularizer values, $\lambda \in \{0.9, 2.0\}$, are illustrated in Fig. \ref{fig:fig3exp}(b)(iii-iv). A dotted line at $x_2 = 53(5)\mu m$ serves as a reference point for subsequent line-cut analyses, as shown in Fig. \ref{fig:fig3exp}(d). The plots pertaining to regularizer selection for the $\mathcal{B}$ method are presented in Fig. \ref{fig:fig3exp}(c).

Upon visual examination, the reconstructions of the $\mathcal{F}_{XY}$ and $\mathcal{F}_Z$ cases exhibit notable similarity, with moderate edge artifacts apparent in the $\mathcal{F}_Z$ reconstruction. This outcome aligns with expectations, as the magnetic field $B_z$ experiences significant decay towards the top and bottom sides of the sample image, resulting in minimal truncation artifacts. However, both Fourier reconstructions display considerable Gaussian noise, evident in the density maps and the total integrated current at $x=0$, measuring $956(7)\mu A$ for $\mathcal{F}_{XY}$ and $1271(4)\mu A$ for $\mathcal{F}_Z$. Both values deviate significantly from the anticipated $700 \mu A$.

In contrast, the Bayesian reconstructions do not exhibit Gaussian noise outside or on the wire, although the $\lambda = 0.9$ reconstruction displays minor aberrations beyond the wire region, which are observed to diminish in the $\lambda = 2$ case. The choice of $\lambda = 0.9$ was determined from the turn in the $l$-curve shown in Fig. \ref{fig:fig3exp}(d)(i), yet the total integrated current at $x = x_2$ reaches $700\mu A$ at $\lambda = 2$ (Fig. \ref{fig:fig3exp}(d)(ii)). While the total current at $x_2$ and the total average current, plotted in the two axes of Fig. \ref{fig:fig3exp}(d), both converge to approximately $700\mu A$ at a higher strength of $\lambda = 4.7$, the reconstruction at such high regularization appears excessively smoothed, lacking the finer details of current flow in the wire. Additionally, the $\lambda = 0.9$ reconstruction reports a total entering current of $772\mu A$---showing significantly less deviation than the Fourier methods---while the $\lambda = 2$ reconstruction shows an entering current of $695(7)\mu A$. This further underscores that the $l$-curve, when combined with the total current, offers a reliable method for selecting the appropriate regularization in Bayesian algorithms for widefield imaging scenarios. The line-cut plots in Fig. \ref{fig:fig3exp}(e) support this analysis by illustrating (1) the edge artifacts in the $\mathcal{F}_Z$, (2) noise in both Fourier images, (3) overestimation at $\lambda = 0.9$, and (4) a noise-free solution at $\lambda = 2$, consistent with the maximum values set by the Fourier methods. 

\section{Conclusions\label{sec:conclusions}}

In conclusion, we present a thorough comparison between Fourier-based and Bayesian reconstruction algorithms using synthetic data with simulated noise models and experimental data obtained from widefield NV diamond magnetometry. Our findings indicate that, when only out-of-plane magnetic field data is available, the Bayesian method outperforms the out-of-plane Fourier method in several regimes. This superiority stems from the Bayesian method's systematic noise rejection and ability to reconstruct the underlying density with high resolution. Conversely, the out-of-plane Fourier method tends to incorporate sensing noise into the reconstruction outside the wire, displaying edge artifacts. In the realm of vector magnetometry, the in-plane Fourier method yields optimal reconstructions under low standoffs and moderate noise conditions. However, the Bayesian algorithm proves advantageous in scenarios involving high standoffs and/or substantial noise, delivering accurate current density magnitudes and sharp resolution. We have outlined a method for selecting the regularizer strength $\lambda$ based on a rough understanding of the typical standoff in an experiment. This approach has resulted in physically accurate reconstructions with correct total currents and minimal aberrations. This work establishes a foundation for applying recently proposed Bayesian algorithms to magnetic fields acquired in a standard widefield NV center quantum diamond microscopy. Additionally, it extensively discusses the limitations associated with existing Fourier methods.

\section{Acknowledgments}

K.S. acknowledges funding from AOARD grant number FA2386-23-1-4012, DST-QuEST grant number DST/ICPS/QuST/Theme-2/2019/Q-58 and DST-SERB Power Grant SPG/2023/000063. K.S. acknowledges support from IITB Nanofabrication facility. S.M. acknowledges funding from I-Hub Chanakya Undergraduate Fellowship. D.A.B. and J.-P.T. acknowledge support by the Australian Research Council (ARC) through grants FT200100073, DP220100178, and DE230100192.

\bibliography{reconsbib}

\onecolumngrid
\renewcommand{\figurename}{Supplementary Figure}
\setcounter{figure}{0}    

\newpage

\section*{Supplementary Material}
First, the simulation methodology is introduced, followed by a discussion of the reconstruction algorithms and the error-estimation metrics employed in the main text. Then, a comprehensive explanation of the regularizer selection in the Bayesian algorithm for the spiral wire case is presented. Lastly, details of the experiments are discussed.

\section{Simulated Magnetic Fields}
Only quasi-two-dimensional current densities are considered, wherein the out-of-plane part of the current is neglected. Hence $\mathbf{j} = (j_x, j_y, 0)$. For computing the magnetic field resulting from a current source, the Biot-Savart reads,
\begin{equation}
    \mathbf{B}(\mathbf{r}) = \frac{\mu_0}{4\pi}\int d^3\mathbf{r}' \frac{\mathbf{j}(\mathbf{r}') \times (\mathbf{r} - \mathbf{r}')}{\mid\mathbf{r} - \mathbf{r}'\mid^3}
\end{equation}
where $\mu_0$ is the vacuum permeability constant. In the Fourier space, this relation can be simplified to \cite{techniques_broadway_vector,roth_using_1989} 
\begin{equation} \label{eq:biosava}
\begin{pmatrix}\mathfrak{B}_x \\ \mathfrak{B}_y \\ \mathfrak{B}_z \end{pmatrix} = \mathcal{M}\begin{pmatrix} \mathfrak{J}_x \\ \mathfrak{J}_y\end{pmatrix} =  \frac{1}{\alpha}\begin{pmatrix}0 & 1 \\ -1 & 0 \\ \iota k_y/k & -\iota k_x/k \end{pmatrix}\begin{pmatrix} \mathfrak{J}_x \\ \mathfrak{J}_y\end{pmatrix}
\end{equation}
where the $\mathfrak{B}$ and $\mathfrak{J}$ denote the Fourier-space magnetic fields and current densities, respectively. The factor $\alpha = 2e^{kz}/\mu_0$ depends on the distance between the sensor and the source plane, $z$, which is the standoff. \\\noindent 
The magnetic field simulations are performed by (i) specifying the (two-dimensional) wire structure and magnitude of the current density map, (ii) applying the Biot-Savart law, and subsequently (iii) adding suitable noise. Aiming to analyze a structure with non-trivial features and highlight nuances of the reconstruction algorithms, we work with a micro-coil wire with multiple sharp turns. The current density across this wire is held constant at $100 A / m$, and the $x$ and $y-$components of the current density are specified piecewise in each arm of the wire. We compute the fields resulting from this wire at a standoff of $z~\mu m$ using a Fourier-space application of the Biot-Savart kernel $\mathcal{M}$. Now, as any experimental data contains noise, we add noise to these simulated images to (i) emulate experimental conditions and (ii) evaluate the noise robustness of the algorithms. The noise is added in two regimes -- (i) Gaussian noise and (ii) scaled noise. The amount of noise under both the regimes is controlled by a single parameter $\eta$, as follows: Given a matrix $A \equiv A(i,j) \in \mathbb{R}^{n \times m}$ we add noise as follows:
 \begin{equation}
        A(i,j) \mapsto A(i,j) + N(i,j),
    \end{equation}
with the two cases being defined as ($[n]$ denotes $\{1,2,\dots n\}$)
\begin{enumerate}
    \item \textit{Gaussian Noise}: For every $i \in [n]$ and $j \in [m]$, we have $N(i,j) \sim \mathcal{N}(0, \sigma^2)$ with $\sigma := \eta \times \max_{i,j} 
    |A(i,j)|$. Thus, $\sigma_{ij} = \sigma~ \forall i\in [n],j\in[m]$.
    \item \textit{Scaled Noise}: We set $N(i,j) \sim \mathcal{N}(0, \sigma_{i,j}^2)$ with $\sigma_{i,j} := \eta \times |A(i,j)|$. 
\end{enumerate}
Note that the Gaussian case adds uncorrelated Gaussian noise of the \textit{same} strength at all pixels. In contrast, scaled noise is more pronounced at pixels with a `large' absolute and less pronounced at pixels with a small value. Both the noise regimes are controlled by the same parameter $\eta$; note that for a fixed $\eta$, the two regimes will be equivalent for a constant image. In the main text, we distinguish between Gaussian and scaled noise by writing $\eta \equiv \eta_{g}$ or $\eta \equiv \eta_{s}$ respectively.\\ 

\section{Reconstruction Methodologies}
When all the components of $\mathbf{B} = (B_x,B_y,B_z)$ are known, say in a vector measurement setup, the problem of reconstructing the $\mathbf{j}$ becomes over-constrained \cite{techniques_broadway_vector} and one can subsequently devise two methods involving a direct Fourier inversion using different components of Eq. \ref{eq:biosava}. The first method uses information obtained from the in-plane measurements $B_{x,y}$, and the second one relies on only the out-of-plane component $B_z$. We refer to them as the the Fourier-XY and the Fourier-Z methods respectively, denote them $\mathcal{F}_{XY}$ and $\mathcal{F}_Z$. We have,

\begin{equation}
(B_x, B_y) \xrightarrow[]{\mathcal{F}_{XY}}  \begin{cases}
\mathfrak{J}_x &:= -\alpha\mathfrak{B}_y, \\ 
\mathfrak{J}_y &:= \alpha\mathfrak{B}_x.
\end{cases}
\end{equation}
and
\begin{equation}
B_z \xrightarrow[]{\mathcal{F}_{Z}}  \begin{cases}
\mathfrak{J}_x &:= \frac{\alpha k_y}{\iota k}\mathfrak{B}_z, \\ 
\mathfrak{J}_y &:= \frac{-\alpha k_x}{\iota k}\mathfrak{B}_z,
\end{cases}
\end{equation}
where, $\mathfrak{B}_i$ ($i=x,y,z$) denotes the $i-$component of the Fourier-space magnetic field, $\mathfrak{J}_k$ ($k = x,y$) denotes the $k-$component of the Fourier-space current density, and $\phi \xrightarrow[]{\mathcal{X}}$ denotes the application of the method $\mathcal{X} \in \{\mathcal{F}_Z, \mathcal{F}_{XY}\}$ on the input magnetic field data $\phi \subset \{B_i\}_i$. Note the $1/k$ pole terms in the $\mathcal{F}_Z$ reconstruction method. These set the stage for the Fourier reconstructions considered in this manuscript.

The Bayesian method works by casting the reconstruction problem as an inference problem and subsequently solving it by mapping it to an optimization procedure. The idea of such methods is to learn the `best fit' current density via maximization of the posterior probability $p(\mathbf{j}|\phi)$, which equates to finding the current densities that are most probable given our magnetic field data. Here $\phi$ denotes the sensed fields. Conservation of current is built-in into the Bayesian algorithm by defining an auxiliary $g$ field and defining the vector current density as $\mathbf{j} = \nabla \times (g\hat{\mathbf{z}})$, which ensures $\nabla\cdot\mathbf{j} = 0$ \cite{clement_reconstruction_2021}. This procedure thus aims to get the most probable current density producing the given field while accounting for the noise in the imaging process.

Usually, an additive noise model is assumed, where the sensed field is given as $\mathbf{B} = \mathcal{M}(g) + \eta$ where $\mathcal{M}$ represents the Biot-Savart operator, and $\eta$ the additive noise. Typically, the additive noise model assumed by the Bayesian algorithm is independent and identically distributed Gaussian noise across all the pixels. We use the Bayes' rule, which reads, $p(g|\phi) \sim p(\phi|g)p(g)$ -- thereby reducing the posterior probability as a product of the \textit{likelihood} $p(\phi|g)$ and the \textit{prior} $p(g)$. Mathematically,
\begin{equation}\label{eq:maxprob}
    g^* = \max_g~p(g|\phi) = \max_g~p(\phi|g)p(g)
\end{equation}
The likelihood is subsequently computed using the additive noise model: $p(\phi|g)$ = $p(\eta = \mathbf{B} - \mathcal{M}(g))$ and the prior $p(g)$ allows for a certain degree of freedom. The form of the noise $\eta$ is assumed to be Gaussian with variance $\sigma^2$, leading to, 
\begin{equation}
    p(\phi|g) = \frac{1}{(2\pi\sigma^2)^{N/2}}\exp{\left(-\frac{||\mathcal{M}(g) - \phi||_2^2}{2\sigma^2}\right)}
\end{equation}
where $N=N_xN_y$ is the number of pixels in the image. The prior $p(g)$ with strength $\lambda$, is used to impose constraints on the reconstruction, typically written as $p(g) \propto \exp{(-\lambda^2\mathcal{L}(g))}$, where $\mathcal{L}$ is some non-negative loss function. This process of imposing a prior is equivalent to a adding regularizer of strength $\lambda$ in the optimization, which attempts to reject the noise. This can be seen by expanding out Eq. \ref{eq:maxprob}: once a $\lambda$ is fixed, the optimal reconstruction is obtained by solving the optimization problem
\begin{equation}
    g_{\lambda}^* := \min_{g}\left( \underbrace{\frac{1}{2}||\mathcal{M}(g) - \phi||_2^2}_{\textbf{Data fitting}} + \underbrace{\lambda^2\mathcal{L}(g)}_{\textbf{Noise rejection}}\right) \label{eq:optim}
\end{equation}
The ratio between the noise-rejecting term and the data-fitting term is now controlled by $\lambda$. \cite{clement_reconstruction_2021, techniques_feldman_regul}. For instance, the total-variation (TV) prior penalizes uncontrolled oscillations in the reconstruction by penalizing derivatives of $\mathbf{j}$. The TV prior is defined as,
\begin{align}
    \mathcal{L}_{TV}(j) &= \iint d^2r (|\partial_xj_y|^2 + |\partial_yj_x|^2) \\ 
    \implies \mathcal{L}_{TV}(g) &= \iint d^2r (|\partial^2_xg|^2 + |\partial^2_yg|^2),   
\end{align}
where the second equation follows from the first because $j = \nabla \cross (g\hat{z})$. We fix the TV prior in our analysis of the Bayesian reconstructions, and acknowledge the use of the \texttt{pysquid} library \cite{clement_reconstruction_2021} for the numerical implementation of the optimization procedures.\\

The results upon application of the three reconstruction algorithms on the spiral wire for the no-noise case of standoff $1\mu m$ are shown in Fig. 1 of the main text. {The Bayesian algorithm is run for either 200 iterations or till the solution is converged upto accuracy $10^{-8}$, whichever occurs first. The Alternating Difference Method of Multipliers (ADMM) method \cite{BoydADMM} is used to solve the optimization problem in Eq.~\ref{eq:optim}. The time for a single run of the algorithm (until convergence or maximum number of iterations) varies with the size of the input magnetic field image, as well as the standoff and noise conditions. For the simulation images of size $55 \cross 55$ pixels, the worst-case time for 200 iterations is found to be $t=16s$. For experimental spiral wire, of size $100 \cross 131$ pixels, the algorithm takes at most $t=17s$ to run for 200 iterations. For the Niobium experiment, featuring magnetic images of size $201\cross 201$ pixels, the time taken to complete 200 iterations is $t=90s$. Lastly, for the Graphene wire images of size $256\cross 256$ pixels, the time for 200 iterations was $t = 121s$.}

\section{Error estimation}
We use two error estimation metrics to analyze the robustness of the methods in varying noise and standoff conditions. First, we define a distance measure using the two-norm as a simple measure of the quality of reconstruction; the two-norm distance between two real matrices $A \equiv A(i,j)$ and $B \equiv B(i,j)$ denoted as $||A-B||_2$, can be written as,
\begin{equation}
   ||A - B||_2:= \left(\sum_{(i,j)}[A(i,j) - B(i,j)]^2\right)^{1/2}
\end{equation}
Moreover, $||A - B||_2 = 0$ iff $A = B$. Now, given the ground truth density $\mathbf{j}$ and the reconstructed density $\hat{\mathbf{j}}$, define, 
\begin{equation}
    D_{L2}(\mathbf{j},\hat{\mathbf{j}}) := \sqrt{||j_x - \hat{j}_x||^2_2 + ||j_y - \hat{j}_y||_2^2}
\end{equation}
We drop the bold-face for brevity and write $D_{L2}(j,\hat{j})$ further. Note that $D_{L2}(j,\hat{j})=0$ iff both the densities have both the $x$ and $y$ components equal. This serves to quantitatively comment on the robustness of an algorithm -- when the noise in the simulation data is ramped up; we analyze the distance as a function of $\eta$ and comment on the robustness using such $D_{L2}$ vs. $\eta$ plots. However, we note that when presented with extremely noisy data, the task is not to match the reconstruction in two-norm; rather, what is of utmost importance is to have a reconstruction qualitatively reflecting the physics in the sample. Formally, two reconstructions might be exactly similar except for one large artifact at a single pixel, which does not negatively affect the quality of the reconstruction. This calls for a distance measure that accounts for perceptual image quality. We thus let our second distance measure be the structured similarity index measure (SSIM), first developed in \cite{wang2004image}. For two matrices $A,B$ of spatial size $L$ this is defined as,
\begin{equation}
    \text{SSIM(A,B)} := \frac{(2\mu_A\mu_B + c_A)(2\sigma_{AB} + c_B)}{(\mu_A^2 + \mu_B^2 + c_A)(\sigma_A^2 + \sigma_B^2 + c_B)}
\end{equation}
where $\mu_X, \sigma_X$ are the mean and standard deviations of the image $X \in \{A,B\}$, $\sigma_{AB}$ is the covariance of $A$ and $B$. The variables $k_1, k_2 \leqq 1$ are constants, which are used to define the stability constants $c_X = (k_XL)^2$ for $X \in \{A,B\}$. As we show in the main text, the SSIM is a better metric for the current reconstruction problem owing to its ability to quantify the key features of the reconstructions, as well as return a bounded score ($0 \leq$ SSIM $\leq 1$).

\section{Regularizer Selection in Bayesian Reconstructions}

\begin{figure*}[t]
\includegraphics[width=0.95\textwidth, height = 11cm]{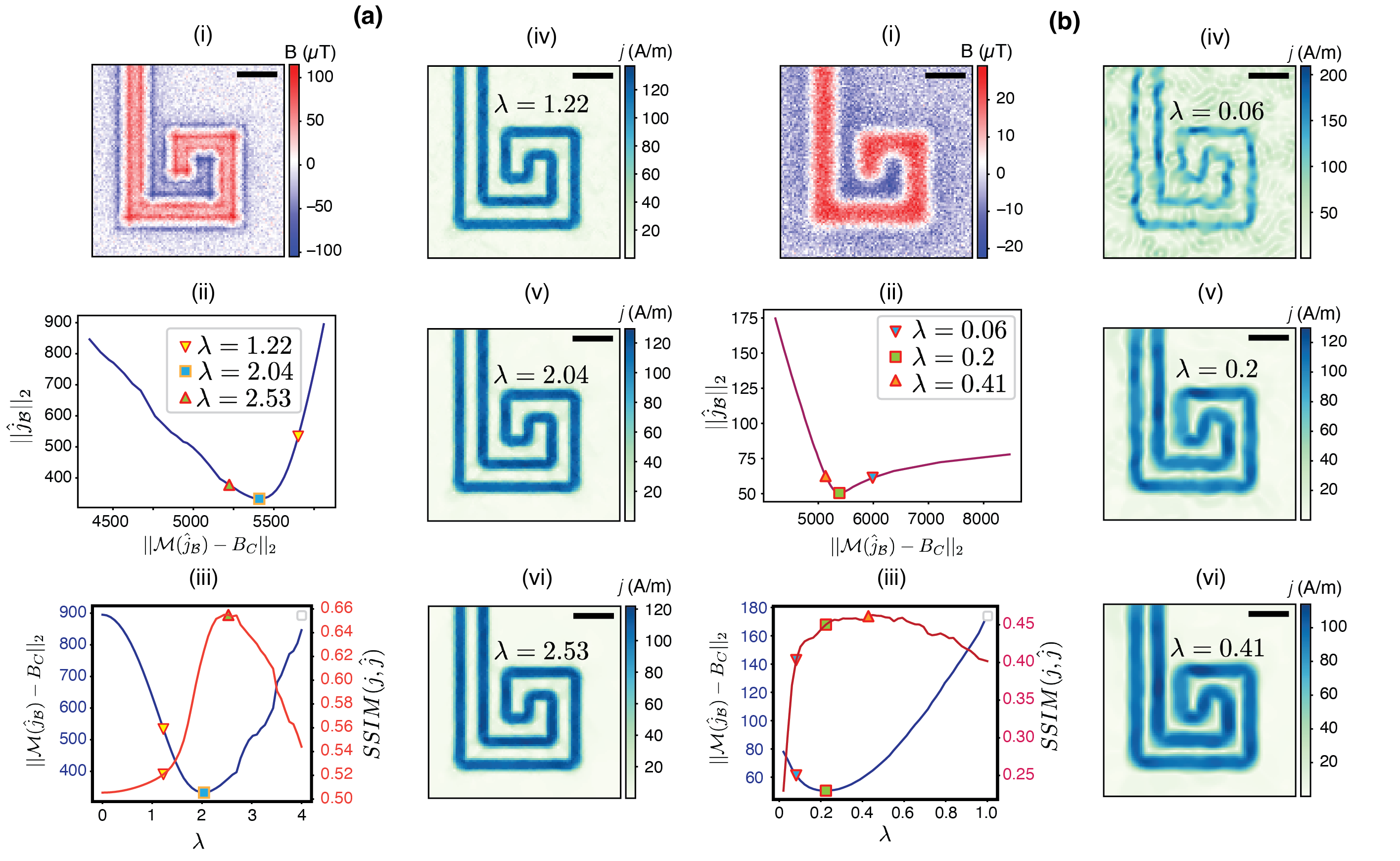}
\caption{Procedure for selecting the regularizer strength for a constant Gaussian noise with strength $\eta_g = 0.5$. (a) Results for the small standoff regime of $z = 0.1\mu m$. (b) Results for the high standoff of $z = 5 \mu m$. Subpanels of (a-b): (i) Noise magnetic field. (ii)  $l$-curve with three regularizer strengths marked for inspection. (iii) The residual distance between the clean and reconstructed magnetic field (left axis) and the SSIM distance between the true and reconstructed densities (right axis). (iv-vi) Reconstructions for three cases of regularizer strengths marked in the $l$-curve.}
\label{fig:fig5}
\centering
\end{figure*}

The regularizer $\lambda$ in the Bayesian method can be used to control the features of the current density image. A small $\lambda \approx 0$ fits to the data, which can include the noise, and will produce exactly the same results as the Fourier inversion in the case of no noise ($\eta = 0$). The ingenuity of the Bayesian methods lies in tuning the $\lambda$, to be able to control the tradeoff between fitting the exact noisy data and rejecting the noise to reconstruct the correct fields. This calls for systematic procedures to choose the optimal regularizer strength $\lambda = \lambda^*$ for a given reconstruction. Previous works on current reconstruction have outlined the discrepancy principle for regularization selection \cite{clement_reconstruction_2021}, which relies on a residual defined as 
\begin{equation}
    r_{\lambda} := \mathcal{M}(\hat{j}_{\lambda}) - B_c
\end{equation}
and outputs $\lambda = \lambda^*$ if the noise in the residual matches with the actual noise of the sensing, i.e., $\text{std}(r_{\lambda}) = \sigma$. The challenge with such approaches lies in the complexity of determining the noise level, denoted as $\sigma$, in a widefield NV experiment. This complexity necessitates a method that is not reliant on $\sigma$. We employ the $l$-curve approach \cite{hansen1999curve,Kindermann_2020,TV_lcurve} in this work for selecting the regularizer strength, establishing its significance as a experimentally-relevant method. This method involves plotting a parameterized curve representing the norm of the residual $||\mathcal{M}(\hat{j}_{\lambda}) - B||_2$ against the norm of the reconstruction $||\hat{j}||_2$. This visualization explores the trade-off between fitting the data and regularizing the solution, revealing a distinct `kink' in the curve that corresponds to the correct regularizer strength. Although this method is originally analytical for Tikhonov regularization \cite{Golub1999TikhonovRA}, where the norm of the reconstruction is regulated, we demonstrate its effectiveness for total variation regularization as well. To support this, we include a plot of the SSIM$(j,\hat{j})$.

The results are shown in Supp.~Fig.\ref{fig:fig5}, for (a) low standoff of $z=0.1\mu m$ and (b) high standoff of $z = 1\mu m$. The panels show (i) the noisy magnetic field, (ii) the $l-$curve, (iii) a plot of the residual norm as well as the SSIM score, and (iv-vi) some reconstructed densities corresponding to $\lambda$-points marked in the $l-$curve. The noise level is kept constant at $\eta_g = 0.5$ for both the standoff regimes.  

The $l$-curve in Supp.~Fig.\ref{fig:fig5}(a) shows a turn at $\lambda = 2.04$, with two additional points marked at $\lambda \in \{1.22,2.53\}$. The value $\lambda = 2.53$ was chosen in line with the plot of SSIM as a function of $\lambda$ as shown in Supp.~Fig.\ref{fig:fig5}(a)(iii), and corresponds to the maximum similarity distance between the reconstructed and true densities. We show the reconstructed densities at these three strengths in Supp.~Fig.\ref{fig:fig5}(a)(iv-vi), and remark that $\lambda = 1.22$ shows noise artifacts inside and outside the wire and is thus before the turn in the $l-$curve. The $\lambda=2.04$ shows no artifacts, but has a higher offset. Instead, the maxima of the SSIM curve at $\lambda = 2.53$ shows a lower offset with no artifacts outside the wire. This demonstrates that the $l-$curve is not exact for TV regularization and should instead be used in supplement with SSIM to choose the regularizer strength.

The outcomes for a high standoff are presented in Supp.~Fig.\ref{fig:fig5}(b). The $l$-curve exhibits a distinct inflection point at $\lambda = 0.2$, with two supplementary points $\lambda \in \{0.06,0.41\}$ marked for reference. However, the SSIM curve displays a noticeable contrast compared to the low standoff scenario. Instead of a peak, there is a plateau, and a reference local maximum is identified on this plateau at $\lambda=0.41$. Additionally, as anticipated, the SSIM distance is lower than in the low standoff case. This indicates that the SSIM curve alone is insufficient for regularizer selection, as all models with $\lambda \in [0.2,0.5]$ yield the same SSIM, rendering it challenging to make a selection. This issue is further elucidated by the reconstructions in Supp.~Fig.\ref{fig:fig5}(b)(iv-vi). The image corresponding to $\lambda = 0.2$, identified by the kink in the $l$-curve, demonstrates the optimal balance between resolving wire features and suppressing noise. In contrast, the $\lambda= 0.06$ image exhibits excessive noise, while the $\lambda = 0.41$ plot shows significant broadening. Consequently, employing the $l$-curve in conjunction with an additional distance metric like SSIM offers a reliable method for selecting the regularizer strength in Bayesian optimization.

We conclude this discussion by highlighting that the $l$-curve was employed due to the absence of information regarding the actual noise level $\sigma$ in an experiment. However, akin to the unknown $\sigma$, the true $j$-map is also unknown, and we only have access to it through numerical simulations in an experimental setting. Consequently, when applying this method to experiments, we substitute the SSIM metric with the \textit{total integrated current}. This choice is motivated by the need to strike a balance between over-fitting and over-smoothing in a conserved quantity, specifically the total current across a cross-section.

\section{Experimental Details}
Experimental magnetic field images have been acquired on a standard wide field-of-view NV microscope, as broadly explained in Ref. \cite{levine2019principles}. 

\subsection{The $\mu$-spiral wire}
For the $\mu$-spiral wire, a CMOS camera setup as described in Ref. \cite{nv_wf_madhur} was used. A schematic is shown in Supp.~Fig.~\ref{FigSI_Nb_graphene}(a) along with a micrograph of the fabricated coil in Supp.~Fig.~\ref{FigSI_Nb_graphene}(b). A single NV resonance profile was selected out of eight Zeeman resonances with high peak-to-peak magnitude for the highest sensitivity, and 2D NV fluorescence maps were acquired as a function of varying microwave frequency for current-off and current-on conditions through the $\mu$-wire. Resonant frequency for each pixel for both current-on and current-off cases were extracted via non-linear fitting of resonance line-shapes. The `three standard deviations away from the mean' criteria was used to remove some noisy or artifact pixels. The difference in the resonant frequency of each NV pixel scaled by the Gyromagnetic ratio ($28MHz/mT$) provided the magnetic field experienced by small ensembles of NV centers within that pixel. Per-pixel spatial resolution in the experimental setup, excluding additional standoff spatial blurring, is $1.33-1.7\mu m$ and depends on the microscope magnification $(23-30)$X. The median per-pixel sensitivity of the experimental setup was $731nT/\sqrt{Hz}$.

\subsection{Niobium wire}

\begin{figure*}[t]
\includegraphics[width=0.6\textwidth]{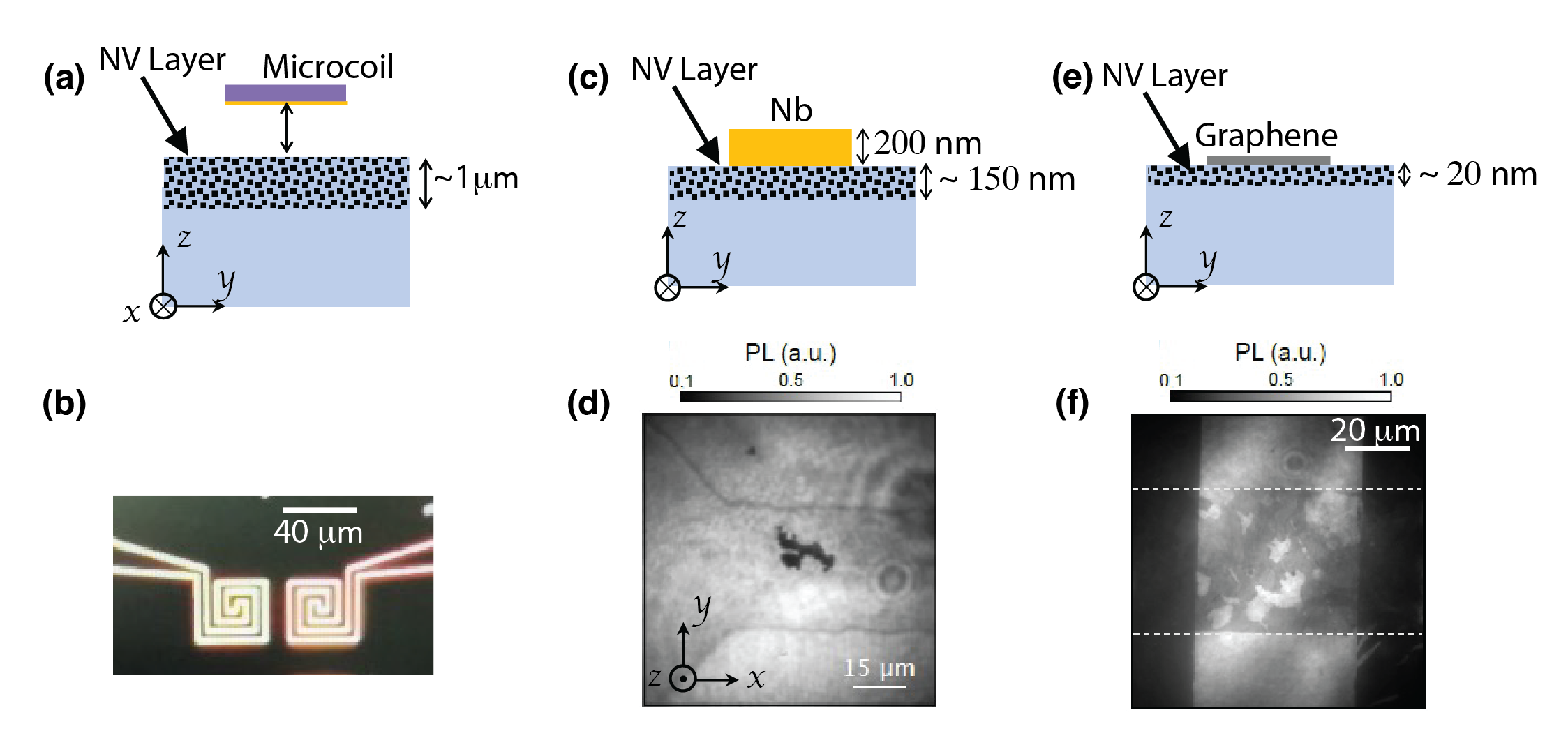}
\caption{Top row (a,c,e): Schematic cross-section of the (a) spiral device, (c) niobium device, and (e) graphene device. Bottom row (b,d,f): Photoluminescence (PL) image of each device corresponding to the same region analysed in the main text.}
\label{FigSI_Nb_graphene}
\centering
\end{figure*}

The niobium (Nb) wire (thickness 200 nm) was fabricated directly on the surface of a diamond which had an NV layer of about 150 nm thickness, see schematic cross-section in Supp.~Fig.~\ref{FigSI_Nb_graphene}(c) and fabrication details in Ref.~\cite{Lillie2020}. The Nb wire is $40\,\mu$m wide in the narrowest section, and widens towards the left of the image analysed in the main text, see corresponding PL image in Supp.~Fig.~\ref{FigSI_Nb_graphene}(d). Vector magnetometry was performed under a small bias magnetic field of $B = 100$\,G, following the procedure described in Ref.~\cite{Tetienne2019}. Measurements were first taken with a dc current of 20 mA passing through the Nb wire, and then again with no current applied. The magnetic field maps analysed in the main text correspond to the difference between the maps obtained with and without current, thus removing any non-current-induced magnetic fields. 

\subsection{Graphene wire}

The graphene wire (monolayer graphene) was fabricated directly on the surface of a diamond which had an NV layer of about 20 nm thickness, see schematic cross-section in Supp.~Fig.~\ref{FigSI_Nb_graphene}(e) and fabrication details in Ref.~\cite{grapheneFET}. The graphene wire is $50\,\mu$m wide, see PL image in Supp.~Fig.~\ref{FigSI_Nb_graphene}(f). Note, there is also a top gate on part of the device, which appears bright in the PL image. This top gate was not used in the measurements analysed in the main text. Vector magnetometry was performed as for the Nb wire, except that here measurements were taken with a positive current of $+700\,\mu$A and then a negative current of $-700\,\mu$A, and the difference (divided by two) between the two maps was used for analysis in the main text.

\end{document}